\documentclass[aps,pre,twocolumn,superscriptaddress,amsmath]{revtex4}

\newcommand{\Sab}{ S_{\alpha, \beta} }
\newcommand{\sab}{ s_{\alpha, \beta} }
\newcommand{\iab}{ \iota_{\alpha, \beta} }
\newcommand{\lab}{ \lambda_{\alpha, \beta} }
\newcommand{\Cab}{ C_{\alpha, \beta} }
\newcommand{\Cba}{ C_{\beta, \alpha} }
\newcommand{\Dab}[1]{{\rm D}_{#1}^{\alpha, \beta}} 
\newcommand{\Mab}[1]{ M_{#1}^{\alpha, \beta}}

\usepackage{graphics}

\begin{document}
\title{A two-parameter generalization of Shannon-Khinchin Axioms\\
and the uniqueness theorem}

\author{Tatsuaki Wada}
\email{wada@ee.ibaraki.ac.jp}
\affiliation{Department of Electrical and Electronic Engineering, 
Ibaraki University, Hitachi, Ibaraki 316-8511, Japan}

\author{Hiroki Suyari}
\email{suyari@faculty.chiba-u.jp, suyari@ieee.org}
\affiliation{Department of Information and Image Sciences, 
Faculty of Engineering, Chiba University, 263-8522, Japan}

\date{\today}

\begin{abstract}
Based on the one-parameter generalization of Shannon-Khinchin (SK)
axioms presented by one of the authors, and utilizing a tree-graphical
representation, 
we have further developed the SK Axioms in accordance with
the two-parameter entropy introduced by Sharma-Taneja, Mittal,
Borges-Roditi, and Kaniadakis-Lissia-Scarfone.
The corresponding unique theorem is proved.
It is shown that the obtained two-parameter Shannon additivity
is a natural consequence from the Leibniz rule of the two-parameter
Chakrabarti-Jagannathan difference operator.
\end{abstract}

%
%

\maketitle

\section{Introduction}
We often encounter complex systems which obey asymptotic power-law distribution
in many fields such as high-energy physics, biophysics, turbulence, scale-free 
networks, economic science and so on. 
In order to explain
the statistical natures of such systems, one of the fundamental approaches is 
a generalization
of statistical mechanics in terms of a suitable
generalization of the Boltzmann-Gibbs-Shannon (BGS) entropy.
Tsallis' nonextensive thermostatistics \cite{Tsallis98,NEXT01,NEXT03,book}
is one of such generalizations. Naudts \cite{Naudts04} has developed
the generalized thermostatistics based on deformed exponential and 
logarithmic functions in general context.\\
In 1975 Sharma and Taneja \cite{Sharma75}, and independently 
Mittal \cite{Mittal75} obtained a two parameter entropy in the field
of information theory by generalizing Chaundy and McLeod's functional
equation which characterizes Shannon' entropy.
In the field of statistical physics, quite recently Kaniadakis, Lissia 
and Scarfone \cite{KaLiSc04, KaLiSc05} have considered 
a differential-functional equation imposed by the MaxEnt principle,
and obtained the two-parameter ($\kappa$ and $r$) entropy,
\begin{equation}
   S_{\kappa, r} = - \sum_i p_i^{1+r} \left(
      \frac{p_i^{\kappa}-p_i^{-\kappa}}{2 \kappa} \right),
  \label{KLS-entropy}
\end{equation}
which is equivalent to the Sharma-Taneja-Mittal entropy.
For the sake of simplicity Boltzmann' constant $k_{B}$ is set to unity
in this paper.
The two-parameter entropy $S_{\kappa, r}$ includes some 
one-parameter generalized entropies which proposed 
by Tsallis \cite{Tsallis88}, by Abe \cite{Abe97} and 
by Kaniadakis \cite{Kaniadakis01} as a special case. For examples,
when $r=\kappa$ and $q=1-2\kappa$, $S_{\kappa, r}$
reduces to Tsallis' entropy
\begin{equation}
   S_q = \frac{1 - \sum_i p_i^{q}}{q-1},
  \label{Tsallis}
\end{equation}
and when $r=0$, $S_{\kappa, r}$ reduces to Kaniadakis'
entropy
\begin{equation}
   S_{\kappa} 
   = \sum_i \frac{p_i^{1+\kappa} - p_i^{1-\kappa}}{2 \kappa}.
  \label{k-entropy}
\end{equation}
Consequently the generalization of thermostatistics
based on the two-parameter entropy provides us a unified framework
of non-extensive thermostatistics. It has been shown
that the two-parameter entropy has some important thermostatistical properties,
such as positivity, continuity, expandability, concavity, Lesche stability, 
and so on \cite{KaLiSc04,KaLiSc05}.
Thermodynamic stability for microcanonical systems described by the 
two-parameter entropy has been studied in Ref. \cite{Scarfone05}.
Scarfone \cite{Scarfone06} has further developed the Legendre structure among 
the generalized thermal quantities in the thermostatistics based on 
the two-parameter entropy $S_{\kappa, r}$.

Abe \cite{Abe97} provided the procedure which generates an entropy functional
from the function
\begin{equation} 
  g(s) \equiv \sum_i p_i^s,
  \label{g}
\end{equation} 
where $p_i$
is a probability of $i$-th event. He observed that the BGS entropy
is obtained by acting the standard derivative on $g(s)$ as
\begin{equation}
    \left[ - \frac{d g(s)}{d s} \right]_{s=1}
       = - \sum_i p_i \ln p_i = S^{\rm BGS},
\end{equation}
whereas Tsallis' entropy is obtained by acting 
Jackson's $q$-derivative (or $q$-difference operator),
\begin{equation}
   {\rm D}_{x}^{q} f(x) \equiv 
        \frac{f(q x)-f(x)}{(q-1) x},
\end{equation}
as follows,
\begin{equation}
   \left[ - {\rm D}_{s}^{q} g(s) \right]_{s=1}
       = \frac{1-\sum_i p_i^q}{q-1}  = S_q.
   \label{1-parameter}
\end{equation}
Johal \cite{Johal98} has established the connection between Tsallis entropy
for a multifractal distribution and Jackson's $q$-derivative.\\
Based on the same procedure as above, Borges and Roditi \cite{Borges98} has 
obtained the two-parameter generalized entropy,
\begin{equation}
      \left[ - {\rm D}_{s}^{\alpha, \beta} g(s) \right]_{s=1} =
        \sum_i \frac{p_i^{\alpha}- p_i^{\beta}}{\beta-\alpha} = \Sab,
     \label{BR-entropy}
\end{equation}
by using the Chakrabarti and Jagannathan (CJ) difference 
operator \cite{Chakrabarti91}
\begin{equation}
   \Dab{x} f(x) \equiv 
        \frac{f(\alpha x)-f(\beta x)}{(\alpha-\beta) x}, 
     \quad \alpha, \beta \in R.
  \label{CJ-derivative}
\end{equation}
The two-parameter CJ difference operator $\Dab{x}$ includes 
Jackson's $q$-derivative 
as a special case in which $\alpha=q$, and $\beta=1$.
The both two-parameter entropies Eqs. \eqref{KLS-entropy} and
\eqref{BR-entropy} are equivalent each other, and they are
related by 
\begin{equation}
 \kappa= \frac{\beta-\alpha}{2}, \quad\textrm{and}
 \quad 1+r= \frac{\alpha+\beta}{2}.
  \label{k-r}
\end{equation}
Note that Eq. \eqref{CJ-derivative} is symmetric under 
the interchange of the two parameters $\alpha \leftrightarrow \beta$.
Consequently the two-parameter entropy $S_{\alpha, \beta}$ 
has the same symmetry. 

On the other hand, it is well known that BGS entropy can be
characterized by the Shannon-Khinchin (SK) axioms \cite{Shannon,Khinchin57}.
During the rapid progress of Tsallis' thermostatistics, the generalized 
SK axioms were proposed by dos Santos \cite{Santos} and by Abe \cite{Abe00}. 
Later, one of the authors \cite{Suyari04} has generalized the SK
axioms for one-parameter generalization of BGS entropy,
and proved the uniqueness theorem rigorously.
To the best of our knowledge, there is no generalization of SK axioms
for either Kaniadakis' entropy $S_{\kappa}$ or 
the two-parameter entropy $\Sab$.
Since $S_{\kappa}$ is a special case of $\Sab$,
it is a natural to generalize the SK axioms
for the two-parameter entropy $\Sab$. This is the main purpose of this paper.
In the next section we review the one-parameter ($q$) generalization of 
SK axioms, among which the key ingredient is the $q$-generalized Shannon 
additivity. In order to develop a two-parameter generalization
of the Shannon additivity, tree-graphical representation
is utilized. In section III we prove the uniqueness theorem associated
with the obtained two-parameter SK axioms. Some examples of a special
case of the two-parameter entropy are presented in section IV.
In section V it is shown that the two-parameter generalized Shannon additivity
is symmetric under the interchange of the two parameters. The relation
with the Leibniz rule of difference (or derivative) operator is discussed.
Final section is devoted to our conclusion. 

\section{One-parameter generalization of Shannon additivity}
We first briefly review the $q$-generalized Shannon-Khinchin 
axioms \cite{Suyari04},
from which the following one-parameter ($q$) generalization of BGS 
entropy is uniquely determined:
\begin{equation}
    S_q(p_1, \dots, p_n) = \frac{1-\sum_{i=1}^{n} p_i^q}{ \phi(q)},
  \label{Sq}
\end{equation}
with $q \in R^{+}$ and $\phi(q)$ satisfies the following properties i)-iv):
\begin{itemize}
  \item[i)] $\phi(q)$ is continuous and has the same sign as $q-1$;
  \item[ii)] $\lim_{q \to 1} \phi(q) = 0$, and $\phi(q) \ne 0$ for $q \ne 0$;
  \item[iii)] there exists an interval $(a, b) \in R^{+}$ such that $a<1<b$
   and $\phi(q)$ is differentiable on the interval $(a,1) \cup (1,b)$;
  \item[iv)] there exists a positive constant $k$ such that 
              $\lim_{q \to 1} \frac{d \phi(q)}{d q} = \frac{1}{k}$.
\end{itemize}
The properties i)-iv) guarantee that Eq. \eqref{Sq}
reduces to BGS entropy in the limit of $q \to 1$. In fact, by 
applying the l'Hopital's rule, we confirm that
\begin{equation}
    \lim_{q \to 1} S_q 
    = \lim_{q \to 1} \frac{-\sum_{i=1}^{n} p_i \ln p_i}{ \frac{d \phi(q)}{dq}}
    = - k \sum_{i=1}^{n} p_i \ln p_i.
\end{equation}
In physics $k$ is Boltzmann' constant $k_{\rm B}$ (recall we set it unity 
in this paper),
and in information theory $k$ is a suitable constant to set
the base of the logarithm, e.g, when $k = 1/ \ln 2$, the base
of the logarithm becomes two.

Let $\Delta_n$ be defined by the $n$-dimensional simplex
\begin{equation}
   \Delta_n \equiv \left\{ (p_1, \dots , p_n) \Big\vert \; p_i \ge 0, 
      \sum_{i=1}^{n} p_i = 1 \right \}.
  \label{Delta}
\end{equation}
The $q$-generalized SK axioms consist of the following four conditions:
\begin{itemize}
  \item {[GSK1]}{\it continuity}:
         $S_q$ is continuous in $\Delta_n$ and $q \in R^{+}$;
  \item {[GSK2]}{\it maximality}:
        for any $n \in N$ and any $(p_1, \dots, p_n) \in \Delta_n$
\begin{equation}
    S_q(p_1, \dots, p_n) \le S_q(\frac{1}{n},\dots ,\frac{1}{n})
\end{equation}
  \item {[GSK3]}{\it generalized Shannon additivity}:
   if
\begin{eqnarray}
  p_{ij} \ge 0, \quad p_i \equiv \sum_{j=1}^{m_i} p_{ij}, \quad
  p(j \vert i) \equiv \frac{p_{ij}}{p_i}, \nonumber \\ 
  \qquad \forall i=1,\dots, n, \quad
  \forall j=1, \dots, m_i 
\end{eqnarray}
 then the following equality holds
  \begin{align}
    S_q(p_{11}, \dots, p_{n m_n}) &= S_q(p_1,\dots ,p_n) \nonumber \\
    + &\sum_{i=1}^n p_i^q S_q \left( p(1 \vert i),\dots ,p(m_i \vert i) \right)
   \label{q-additivity}
  \end{align}
  \item {[GSK4]} {\it expandability}:
      \begin{equation}
        S_q(p_1, \dots, p_n, 0) = S_q(p_1,\dots ,p_n).
      \end{equation}
\end{itemize}
Note that when $q=1$ the above axioms [GSK1]-[GSK4] reduce
to the original SK axioms \cite{Khinchin57}, respectively.   

\begin{figure}[h]
\begin{center}
\begin{picture}(180,60)
    \put(  0, 26){\circle*{3}}
    \put(  0, 26){\line(2,1){52}}
    \put(  0, 26){\line(2,-1){52}}
    \put(  0, 26){\line(1,0){52}}
    \put( 30, 48){\makebox(0,0)[h]{$p_{11}$}}
    \put( 30, 32){\makebox(0,0)[h]{$p_{21}$}}
    \put( 30,  4){\makebox(0,0)[h]{$p_{22}$}}

    \put( 70, 26){\makebox(0,0)[h]{$=$}}

    \put( 90, 26){\circle*{3}}
    \put( 90, 26){\line(2,1){52}}
    \put(116, 39){\circle*{3}}
    \put( 90, 26){\line(2,-1){52}}
    \put(116, 13){\circle*{3}}
    \put(116, 13){\line(2,1){26}}
    \put(102, 38){\makebox(0,0)[h]{$p_1$}}
    \put(102, 14){\makebox(0,0)[h]{$p_2$}}
    \put(130, 56){\makebox(0,0)[h]{$p(1 \vert 1)$}}
    \put(130, 30){\makebox(0,0)[h]{$p(1 \vert 2)$}}
    \put(130, -4){\makebox(0,0)[h]{$p(2 \vert 2)$}}
  \end{picture}
\end{center}
  \caption{A graphical representation of a set of two successive sub-events
    and associated probabilities Eq. \eqref{prob}
    for $n=2$ and $m_1=1, m_2=2$.}
  \label{prob-graph}
\end{figure}
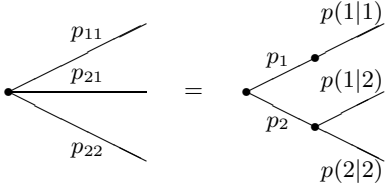

\begin{figure}[h]
\begin{center}
\begin{picture}(260,120)
   \thicklines
    \put(  0, 90){\circle*{3}}
    \put(  0, 90){\line(2,1){40}}
    \put(  0, 90){\line(1,0){40}}
    \put(  0, 90){\line(2,-1){40}}    

    \put( 70, 110){\makebox(0,0)[h]{$ s_q(p_{11}) $}}
    \put( 70,  90){\makebox(0,0)[h]{$ s_q(p_{21}) $}}
    \put( 70,  70){\makebox(0,0)[h]{$ s_q(p_{22}) $}}
    
    \put(100,  90){\makebox(0,0)[h]{$=$}}

    \thinlines
    \put( 10, 20){\circle*{3}}
    \put( 10, 20){\line(2,1){20}}
    \put( 30, 30){\circle*{3}}
    \put( 10, 20){\line(2,-1){20}}    
    \put( 30, 10){\circle*{3}}

    \put( 84, 40){\makebox(0,0)[h]{$p_1^{q} s_q(p(1 \vert 1))$}}
    \put( 84, 20){\makebox(0,0)[h]{$p_2^{q} s_q(p(1 \vert 2))$}}
    \put( 84,  0){\makebox(0,0)[h]{$p_2^{q} s_q(p(2 \vert 2)) $}}
    \thicklines
    \put( 30, 30){\line(2,1){20}}
    \put( 30, 10){\line(2,-1){20}}
    \put( 30, 10){\line(2,1){20}}

    \put(128, 20){\makebox(0,0)[h]{$+$}}

    \put(140, 20){\circle*{3}}
    \put(140, 20){\line(2,1){20}}
    \put(160, 30){\circle*{3}}
    \put(140, 20){\line(2,-1){20}}    
    \put(160, 10){\circle*{3}}

    \put(214, 40){\makebox(0,0)[h]{$ s_q(p_1) p(1 \vert 1) $}}
    \put(214, 20){\makebox(0,0)[h]{$ s_q(p_2) p(1 \vert 2) $}}
    \put(214,  0){\makebox(0,0)[h]{$ s_q(p_2) p(2 \vert 2) $}}
    \thinlines
    \put(160, 30){\line(2,1){20}}
    \put(160, 10){\line(2,-1){20}}
    \put(160, 10){\line(2,1){20}}
\end{picture}
\end{center}
\caption{A tree graphical representation for the specific example
of the q-generalized Shannon additivity Eq. \ref{q-Shannon-add}.}
 \label{Sq-graph}
\end{figure}
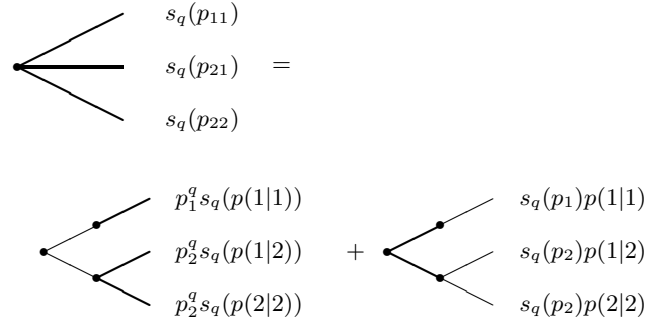

\begin{figure}[h]
\begin{center}
\begin{picture}(260,120)
   \thicklines
    \put(  0, 90){\circle*{3}}
    \put(  0, 90){\line(2,1){40}}
    \put(  0, 90){\line(1,0){40}}
    \put(  0, 90){\line(2,-1){40}}    

    \put( 70, 110){\makebox(0,0)[h]{$ \sab(p_{11}) $}}
    \put( 70,  90){\makebox(0,0)[h]{$ \sab(p_{21}) $}}
    \put( 70,  70){\makebox(0,0)[h]{$ \sab(p_{22}) $}}
    
    \put(100,  90){\makebox(0,0)[h]{$=$}}

    \thinlines
    \put( 10, 20){\circle*{3}}
    \put( 10, 20){\line(2,1){20}}
    \put( 30, 30){\circle*{3}}
    \put( 10, 20){\line(2,-1){20}}    
    \put( 30, 10){\circle*{3}}

    \put( 84, 40){\makebox(0,0)[h]{$p_1^{\alpha} \sab(p(1 \vert 1))$}}
    \put( 84, 20){\makebox(0,0)[h]{$p_2^{\alpha} \sab(p(1 \vert 2))$}}
    \put( 84,  0){\makebox(0,0)[h]{$p_2^{\alpha}\sab(p(2 \vert 2)) $}}
    \thicklines
    \put( 30, 30){\line(2,1){20}}
    \put( 30, 10){\line(2,-1){20}}
    \put( 30, 10){\line(2,1){20}}

    \put(128, 20){\makebox(0,0)[h]{$+$}}

    \put(140, 20){\circle*{3}}
    \put(140, 20){\line(2,1){20}}
    \put(160, 30){\circle*{3}}
    \put(140, 20){\line(2,-1){20}}    
    \put(160, 10){\circle*{3}}

    \put(214, 40){\makebox(0,0)[h]{$ \sab(p_1) p^{\beta}(1 \vert 1) $}}
    \put(214, 20){\makebox(0,0)[h]{$ \sab(p_2) p^{\beta}(1 \vert 2) $}}
    \put(214,  0){\makebox(0,0)[h]{$ \sab(p_2) p^{\beta}(2 \vert 2) $}}
    \thinlines
    \put(160, 30){\line(2,1){20}}
    \put(160, 10){\line(2,-1){20}}
    \put(160, 10){\line(2,1){20}}
\end{picture}
\end{center}
\caption{A tree graphical representation of the two-parameter
generalization of Shannon additivity for the simple case
described in Fig. \ref{prob-graph}}
 \label{graph-rep}
\end{figure}
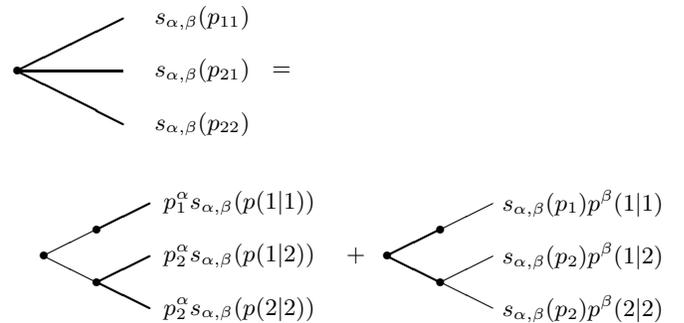

Shannon \cite{Shannon} discussed the synthesizing rule of an entropy
with a tree-graphical representation.
Let us now consider a further generalization of the axiom [GSK3] by
utilizing the similar tree-graphical representation. 
Suppose we have a set of possible events (or choices), and let us
divide each event (choice) into two successive sub-events (choices).
Any joint probability of two successive sub-events can be expressed
as
\begin{align} 
   p_{ij} = p_i \; p(j \vert i),
   \label{prob} 
\end{align}
where $p_i (i=1,\dots,n)$ is a probability 
of $i$-th sub-event and $p(j \vert i)$ a conditional probability, i.e.,
a probability of the $j$-th sub-event ($j=1,\dots, m_i$) after the
$i$-th sub-event occurred.
More specifically, let us consider the following simple case in which
$n=2$ and $m_1=1, m_2=2$. Each probability of any event is graphically
represented by a thin line as shown in Fig \ref{prob-graph}.
Let $S_q$ of Eq. \eqref{Sq} be expressed as the following trace-form
\begin{equation}
   S_q(p_1,\dots,p_n) = \sum_{i=1}^n s_q(p_i),
\end{equation}
where $s_q(p_i) = (p_i - p_i^q)/\phi(q)$. 
Then, for this simple case in Fig. \ref{prob-graph},
Eq. \eqref{q-additivity} in the axiom [GSK3] 
becomes
\begin{align}
    \sum_{i=1}^{2} \sum_{j=1}^{m_i} \; s_q(p_{ij}) =&
    \sum_{i=1}^2 \sum_{j=1}^{m_i} 
           p_i^q  \; s_q \left( p(j \vert i) \right) \nonumber \\
    &+ \sum_{i=1}^2 \sum_{j=1}^{m_i} s_q(p_i) \; p(j \vert i),
    \label{q-Shannon-add}
  \end{align}
where $m_1=1$ and $m_2=2$.
This can be graphically represented
as Fig \ref{Sq-graph}. A thick line represents the $s_q(r)$ of 
a probability $r$ of the corresponding line, 
where $r$ is $p_{ij}$, $p_i$ or $p(j \vert i)$ depending on the line. 
A thin line represents
a weight factor, which is either $p_i^q$ for $i$-th sub-event
or $p(j \vert i)$ for $j$-th sub-event).
Summation over indices is represented by a node in each tree graph.
Note that the weight factor for a first sub-event of successive two 
sub-events is $p_i^q$ whereas the weight factor for a second sub-event
is $p(j \vert i)$.
A natural extension of the one-parameter (q) generalized Shannon
additivity in the axiom [GSK3] to the two-parameter entropy is then to
attribute two different weighs to first and second sub-events, respectively.  
Hereafter we consider a generalized trace-form entropy
\begin{equation}
   \Sab[p_i] = \sum_i \sab(p_i),
   \label{trace-form}
\end{equation}
depending on the two real-parameter $\alpha$ and $\beta$.
Consequently a two-parameter generalization of Shannon additivity 
for the above simple example can be expressed as 
\begin{align}
    \sum_{i=1}^{2} \sum_{j=1}^{m_i} s_q(p_{ij}) =&
    \sum_{i=1}^2 \sum_{j=1}^{m_i} 
           p_i^{\alpha} \; s_q \left( p(j \vert i) \right) \nonumber \\
    &+ \sum_{i=1}^2 \sum_{j=1}^{m_i} s_q(p_i) \; p(j \vert i)^{\beta}.
    \label{ab-Shannon-add}
  \end{align}
Fig \ref{graph-rep} is the graphical representation of 
Eq. \eqref{ab-Shannon-add}.

\section{Two-parameter generalizations of Shannon-Khinchin axioms
and the uniqueness theorem}
Now we propose the two-parameter generalization
of the SK axioms, and prove the unique theorem.

\noindent \textbf{Theorem}: 
Let $\Delta_n$ be an $n$-dimensional 
simplex defined by Eq. \eqref{Delta}.
For a generalized trace-form entropy of Eq. \eqref{trace-form},
the following two-parameter generalized axioms [TGSK1]-[TGSK4] 
determine the function $\Sab: \Delta_n \to R^{+}$ such that
\begin{equation}
   \Sab(p_1, \dots, p_n) 
      = \sum_{i=1}^n \frac{p_i^{\alpha}-p_i^{\beta}}{\Cab},
   \label{Sab}
\end{equation}
where $\alpha$ and $\beta$ are real parameters restricted
within the regions:
\begin{align}
  \Big\{ (\alpha, \beta) \in R^2 &\vert \;
           \alpha \ge 1, \beta \le 1  \Big\} \nonumber\\
  \Big\{ (\alpha, \beta) \in R^2 &\vert \; 
           \alpha \le 1, \beta \ge 1 \Big\}
\nonumber \\
\textrm{ but except } (\alpha, \beta)&=(1,0) \textrm{ and } (0, 1).
  \label{ab-regions}
\end{align}
and $\Cab$ satisfies 
the following properties I)-IV)
\begin{itemize}
  \item[I)] $\Cab$ is continuous w.r.t. $\alpha$ and $\beta$, 
     and has the same sign as $\beta-\alpha$.
     Consequently $\Cab$ is antisymmetric under the interchange of $\alpha$
  and $\beta$, i.e., $\Cba = -\Cab$;
  \item[II)] $\lim_{\alpha \to \beta} \Cab  = 0$, and 
         $\Cab \ne 0$ for $\alpha \ne \beta$;
  \item[III)] there exists an interval $(a, b) \in R$ such that $\Cab$ is 
     differentiable w.r.t. both $\alpha$ and $\beta$ on the interval 
     $(a, 1) \cup (1,b)$;
  \item[IV)] there exists a positive constant $k$ such that
 $\lim_{\alpha \to 1} \frac{d \Cab}{d \alpha} = -\frac{1}{k}$, and
            $\lim_{\beta \to 1} \frac{d \Cab}{d \beta} = \frac{1}{k}$.
\end{itemize} 

\begin{itemize}
  \item {[TGSK1]}{\it continuity}:
         $\Sab$ is continuous in $\Delta_n$;
  \item {[TGSK2]}{\it maximality}:
        for any $n \in N$ and any $(p_1, \dots, p_n) \in \Delta_n$
\begin{equation}
    \Sab(p_1, \dots, p_n) \le \Sab(\frac{1}{n},\dots ,\frac{1}{n})
\end{equation}
  \item {[TGSK3]} {\it two-parameter generalized Shannon additivity}:
\begin{eqnarray}
 \sum_{i=1}^{n} \sum_{j=1}^{m_i} \sab(p_{ij})
  &=& \sum_{i=1}^{n} p_i^{\alpha} \sum_{j=1}^{m_i}
          \sab\left(p(j \vert i) \right)
\nonumber \\
   && + \sum_{i=1}^{n} \sab(p_i)
         \sum_{j=1}^{m_i} p(j \vert i)^{\beta}.
  \label{TGSK3}
\end{eqnarray}
  \item {[TGSK4]} {\it expandability}:
      \begin{equation}
        \Sab(p_1, \dots, p_n, 0) = \Sab(p_1,\dots ,p_n).
      \end{equation}
\end{itemize}

\noindent \textbf{Proof:}
First we consider the special case as same as Eq. (20) of Ref. \cite{Suyari04},
i.e., $\forall m_i = m$ and $p_{ij} = 1/(nm)$.
Then Eq. \eqref{TGSK3} can be written as
\begin{align}
  n m \sab(\frac{1}{nm})
 &= \sum_{i=1}^{n} \frac{1}{n^{\alpha}} 
          m \sab(\frac{1}{m})
   + \sum_{i=1}^{n} \sab(\frac{1}{n}) \sum_{j=1}^{m} \frac{1}{m^{\beta}}
\nonumber \\
  =& n^{1-\alpha} m \sab(\frac{1}{m})
   + n m^{1-\beta} \sab(\frac{1}{n})
  \label{simplified}
\end{align}
Let $\lab(n)$ be defined by
\begin{equation}
     \lab(n) \equiv -\frac{1}{n} \; \sab(n),
  \label{def-lab}
\end{equation}
then Eq. \eqref{simplified} becomes
\begin{equation}
     \lab(\frac{1}{n m}) = n^{1-\alpha} \lab(\frac{1}{m}) 
            + m^{1-\beta} \lab(\frac{1}{n}).
\end{equation}
Exchanging the variables $m$ and $n$, we have
\begin{align}
  n^{1-\alpha} \lab(\frac{1}{m}) +& m^{1-\beta} \lab(\frac{1}{n}) \nonumber \\
   &= m^{1-\alpha} \lab(\frac{1}{n}) + n^{1-\beta} \lab(\frac{1}{m}).
\end{align}
The variable $m$ and $n$ are separated as
\begin{align}
     \frac{n^{1-\beta} - n^{1-\alpha}}{\lab(\frac{1}{n})} =
     \frac{m^{1-\beta} - m^{1-\alpha}}{\lab(\frac{1}{m})} = \Cab,
\end{align}
where $\Cab$ is a constant depending on $\alpha$ and $\beta$.
We thus find
\begin{equation}
     \lab(n) = \frac{n^{\beta-1} - n^{\alpha-1}}{\Cab}.
   \label{lab}
\end{equation}
Next let us take $p_{ij}$ as
\begin{equation}
  p_{ij} = \frac{1}{\sum_{r=1}^{n} m_r},
\end{equation}
for all $i$ and $j$, then
\begin{equation}
  p_i = \sum_{j=1}^{m_i} p_{ij} = \frac{m_i}{\sum_{r=1}^{n} m_r},
\label{pi}
\; \textrm{and} \;
p(j \vert i) = \frac{p_{ij}}{p_i} = \frac{1}{m_i}.
\end{equation}
Eq. \eqref{TGSK3} becomes
\begin{align}
 \sum_{i=1}^{n} \sum_{j=1}^{m_i} \sab\left(\frac{1}{\sum_{r=1}^{n} m_r} \right)
  &=
\sum_{i=1}^{n} p_i^{\alpha} \sum_{j=1}^{m_i}
          \sab\left( \frac{1}{m_i} \right)  \nonumber \\
   +& \sum_{i=1}^{n} \sab(p_i)
         \sum_{j=1}^{m_i} \left( \frac{1}{m_i} \right)^{\beta} 
\end{align}
By utilizing Eqs (\ref{def-lab}) and (\ref{lab}) we have
\begin{align}
 \sum_{i=1}^{n} \sab(p_i) m_i^{1-\beta} &=
  \sum_{i=1}^{n} p_i^{\alpha} \lab\left( \frac{1}{m_i} \right)
    -\lab\left(\frac{1}{\sum_{r=1}^{n} m_r} \right)
\nonumber \\
 = \frac{1}{\Cab} & \left( 
       \sum_{i=1}^{n} p_i^{\alpha} m_i^{1-\beta}-
      \left(\sum_{r=1}^{n} m_r \right)^{1-\beta} \right) 
\nonumber \\
  + \frac{1}{\Cab} & \left( \left(\sum_{r=1}^{n} m_r \right)^{1-\alpha}
    -\sum_{i=1}^{n} p_i^{\alpha} m_i^{1-\alpha} \right)
  \label{eq}
\end{align}
From Eq. (\ref{pi}) it follows
\begin{equation}
    \sum_{i=1}^{n} p_i^{t} \; m_i^{1-t} = 
         \left(\sum_{r=1}^{n} m_r \right)^{1-t},
\end{equation}
with any real number $t$, 
then Eq. \eqref{eq} becomes
\begin{eqnarray}
 \sum_{i=1}^{n} \sab(p_i) \; m_i^{1-\beta} =
 \sum_{i=1}^{n} \left(
        \frac{p_i^{\alpha} -p_i^{\beta}}{\Cab} 
         \right) m_i^{1-\beta}.
\end{eqnarray}
Since we can set $m_i$ arbitrary, by setting $m_i=1$ we finally obtain
\begin{eqnarray}
 \Sab(p_1, \dots, p_n) = \sum_i \sab(p_i) =
    \sum_i \frac{p_i^{\alpha} -p_i^{\beta}}{\Cab}.
\end{eqnarray}

Now we show that $\alpha$ and $\beta$ are in the regions 
of Eq. \eqref{ab-regions} in order to the $\Sab$ is definite concave,
i.e., the second derivative of $\Sab$ w.r.t. the $p_i$ should be negative,
\begin{align}
  \frac{d^2 \Sab}{d p_i^2} &= 
     \frac{\alpha (\alpha-1) p_i^{\alpha-2} - \beta (\beta-1) p_i^{\beta-2}}
    {\Cab} < 0.
  \label{Sab''}
\end{align}
From the property I) we see $(\beta-\alpha) \Cab$ is always positive.
Then the sign of the numerator multiplied by $\beta-\alpha$ should
be negative,
i.e., 
\begin{align}
  (\beta-\alpha) 
     \left\{ \alpha (\alpha-1) p_i^{\alpha-2} - \beta (\beta-1) p_i^{\beta-2}
   \right\} < 0.
  \label{cond-concave}
\end{align}
Let us first consider a simple case in which one of the terms 
in the curly bracket of Eq. \eqref{cond-concave} is vanish.
When $\beta=1$, the condition becomes
\begin{align}
     -\alpha (\alpha-1)^2 p_i^{\alpha-2} < 0.
\end{align}
Since $p_i$ is positive, then $\alpha > 0$.\\
When $\beta=0$, the condition becomes
\begin{align}
     -\alpha^2 (\alpha-1) p_i^{\alpha-2} < 0,
\end{align}
then $\alpha>1$.\\
When $0<\beta<1$, the second term in the curly bracket
is positive. Then the condition becomes $\beta-\alpha < 0$
and $\alpha(\alpha-1)>0$. This is satisfied with
$\alpha>1$.
The rest regions are obtained similar way because the
condition of Eq. \ref{cond-concave} is symmetric under
the interchange of $\alpha$ and $\beta$.
\begin{figure}[h]
\begin{center}
  \resizebox{50mm}{!}{\includegraphics{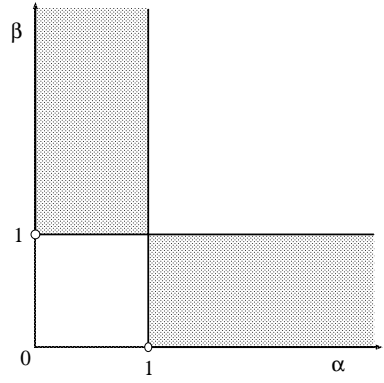}}
\end{center}
\caption{The parameter regions in Eq. \ref{ab-regions}
of $\alpha$ and $\beta$
in which the two-parameter entropy is definitely concave}
 \label{ab-graph}
\end{figure}

The properties II)-IV) are needed in order to the $\Sab$ reduces
to the BGS entropy in the limit of $\alpha, \beta \to 1$.
In fact by applying l'Hopital's rule, we confirm that
\begin{align}
    \lim_{\alpha,\beta \to 1} \Sab 
    &= \lim_{\alpha \to 1} 
          \frac{-\sum_{i=1}^{n} p_i \ln p_i}{ \frac{d \Cab}{d\beta}}
\nonumber\\
    &= \lim_{\beta \to 1} 
          \frac{\sum_{i=1}^{n} p_i \ln p_i}{ \frac{d \Cab}{d\alpha}}
    = - k \sum_{i=1}^{n} p_i \ln p_i.
\end{align}

\section{Some examples of a special case}
For the simplest case in which $\Cab = \beta-\alpha$,
we see that
\begin{equation}
     \lab(x) = \frac{x^{\beta-1} - x^{\alpha-1}}{\beta-\alpha},
     \qquad (x>0).
\end{equation}
Recalling the relations \eqref{k-r} between the entropic parameters 
$(\alpha, \beta)$ and $(\kappa, r)$, we see that $\lab(x)$ is nothing but 
the two-parameter deformed logarithmic function,
\begin{equation}
     \ln_{\{\kappa, r\}}(x) \equiv 
     \frac{x^{r+\kappa} - x^{r-\kappa}}{2 \kappa},
\end{equation}
which is introduced in Ref. \cite{KaLiSc04}.
When $\alpha=1-\kappa$ and $\beta=1+\kappa$, the deformed logarithmic
function reduces to $\kappa$-logarithmic function proposed by Kaniadakis.
\begin{equation}
 \lab(n) \to \frac{n^{\kappa}-n^{-\kappa}}{2 \kappa} = \ln_{\{ \kappa \}} n.
\end{equation}
The entropy $\Sab$ reduces to Kaniadakis' entropy Eq. \eqref{k-entropy}.\\
When $\alpha=q$ and $\beta=1$, it reduces to Tsallis' $q$-logarithmic function
but $q$ replaced with $2-q$
\begin{equation}
 \lab(n) \to \frac{n^{q-1}-1}{q-1} = \ln_{2-q} (n).
\end{equation}
Accordingly $\Sab$ reduces to Tsallis' entropy Eq. \eqref{Tsallis}.
More details on the two-parameter deformed logarithms and
entropies, please refer to Ref. \cite{KaLiSc04}.

Another example is Harvda-Charvat \cite{Havrda} or Dar\'oczi \cite{Daroczi} 
entropy,
\begin{equation}
   S_q^{\rm HCD} = \frac{1-\sum_i p_i^{q}}{1-2^{1-q}},
\end{equation}
which corresponds to the case $\Cab=1-2^{1-\alpha}$,
$\alpha=q, \beta=1$ and $k=1/\ln2$.

\section{on the two-parameter generalized Shannon additivity}
Since $\Cba=-\Cab$, it is obvious from Eq. \eqref{Sab} that $\Sab$ (and $\sab$) 
is symmetric under the interchange of the two-parameter $\alpha$ and $\beta$. 
Then the two-parameter generalized Shannon 
additivity Eq. \eqref{TGSK3} also must hold if $\alpha$ and $\beta$ are
interchanged each other,
\begin{eqnarray}
 \sum_{i=1}^{n} \sum_{j=1}^{m_i} \sab(p_{ij})
  &=& \sum_{i=1}^{n} p_i^{\beta} \sum_{j=1}^{m_i}
          \sab\left(p(j \vert i) \right)
\nonumber \\
   && + \sum_{i=1}^{n} \sab(p_i)
         \sum_{j=1}^{m_i} p(j \vert i)^{\alpha}.
  \label{TGSK3'}
\end{eqnarray}
Then by adding the both sides of Eqs. \eqref{TGSK3} and that of \eqref{TGSK3'}
(and dividing by $2$) we obtain the symmetric form
\begin{align}
 \sum_{i=1}^{n} \sum_{j=1}^{m_i} \sab(p_{ij})
  &= \sum_{i=1}^{n} \iab(p_i) \sum_{j=1}^{m_i}
          \sab\left(p(j \vert i) \right)
\nonumber \\
   +& \sum_{i=1}^{n} \sab(p_i)
         \sum_{j=1}^{m_i} \iab(p(j \vert i)),
  \label{TGSK3-sym}
\end{align}
where we introduced the function
\begin{equation}
 \iab(x) \equiv \frac{x^{\alpha} + x^{\beta}}{2}.
\end{equation}
Recall Eq. \eqref{BR-entropy} in which the two-parameter entropy 
$\Sab$ (and $\sab(p_i))$ is obtained by acting the CJ difference operator 
on $g(s)$ of Eq. \eqref{g}.
Similarly the function $\iab(p_i)$ is obtained from $g(s)$ by acting
the average operator $\Mab{x}$ associated with CJ difference operator 
\begin{equation} 
  \Mab{x} f(x) \equiv \frac{f(\alpha x) + f(\beta x)}{2},
\end{equation}
as follows,
\begin{equation}
    \left[ \Mab{s} g(s) \right]_{s=1} =
        \sum_i \frac{p_i^{\alpha}+ p_i^{\beta}}{2}
       = \sum_i \iab(p_i).
     \label{Iab}
\end{equation}
From the relations in Eq. \eqref{k-r} we see that this is same as the function
\begin{equation}
   {\cal I}_{\kappa,r} \equiv \sum_i p_i^{r+1} 
      \left( \frac{p_i^{\kappa}+ p_i^{-\kappa}}{2} \right),
\end{equation}
which is introduced in Ref. \cite{Scarfone06}, and an important quantity
relating the two-parameter generalized entropy of Eq. \eqref{KLS-entropy}, 
free-energy, partition function, and other thermodynamical quantities.

With the help of the average operator $\Mab{x}$,
the Leibniz rule of the CJ difference operator can be written 
in the symmetric form as
\begin{align}
   \Dab{x} \Big( f(x) g(x) \Big)  
      =& \Big( \Dab{x} f(x) \Big) \Big( \Mab{x} g(x) \Big) \nonumber \\
       &+ \Big( \Mab{x} f(x) \Big) \Big( \Dab{x} g(x) \Big).
  \label{Leibniz_rule}
\end{align}
Then we observe that the two-parameter generalized Shannon 
additivity \eqref{TGSK3-sym} is readily obtained by acting $\Dab{s}$
on $\sum_{i}\sum_j p_{ij}^s = \sum_{i} p_{i}^s \sum_j p(j \vert i)^s$,
as can be seen from the relation
\begin{align}
    \sum_{i, j} \left[ \Dab{s} p_{ij}^s \right]_{s=1} &= 
     \sum_i \left[ \Dab{s} p_i^s \right]_{s=1} 
       \sum_j \left[ \Mab{s} p(j \vert i)^s \right]_{s=1}
        \nonumber \\
     +&\sum_i \left[ \Mab{s} p_i^s \right]_{s=1} 
       \sum_j \left[ \Dab{s} p(j \vert i)^s \right]_{s=1}.
\end{align}
Thus we see that the two-parameter Shannon additivity
is a natural consequence of the Leibniz rule of the CJ difference operator.
In other words, the Shannon additivity associated with an entropy is
the consequence of the Leibniz rule of the corresponding difference (or
derivative) operator which generates the entropy.

Finally let us comment on the number of the parameters
for generalizing the BGS entropy.
One may wonder whether a generalization to more than two parameters
is possible or not. We can answer to this question as follows.
Recall that a parameter generalization
of the BGS entropy is obtained by acting a {\it first-order} difference 
operator on the function $g(s)$, e.g., Eq. \eqref{1-parameter} for
the one-paramere entropy $S_q$ and Eq. \eqref{BR-entropy} for the two-parameter
entropy $\Sab$. Since any first-order difference operator is defined by
the difference of the functions at two points ( 
e.g., Eq. \eqref{CJ-derivative} for $\Dab{x}$), such a generalization of
the BGS entropy is up to two parameters.

\section{Conclusion}
Based on the one-parameter generalized SK axioms \cite{Suyari04} proposed
by one of the authors,
we have further developed the two-parameter generalization of the SK axioms
in accordance with the two-parameter entropy introduced 
by Sharma-Taneja \cite{Sharma75}, 
Mittal \cite{Mittal75}, Borges-Roditi \cite{Borges98}, and 
Kaniadakis-Lissia-Scarfone \cite{KaLiSc05}, 
and proved the corresponding uniqueness theorem.
The Shannon additivity, which is a key ingredient of the SK axioms,
is generalized by considering the tree-graphical representation. 
We have obtained the symmetric form of the two-parameter generalized
Shannon additivity, and shown the relation with the Leibniz rule
of the CJ difference operator.


\end{document}